\documentclass[prd,onecolumn,showpacs,superscriptaddress,floatfix,nofootinbib]{revtex4}
\usepackage{graphicx,psfrag,amsmath,amssymb,amsfonts,bbm,latexsym,color,dcolumn,bm}

\begin{document}

\title{Muonic hydrogen as a quantum gravimeter}

\author{Roberto Onofrio}
\email{onofrior@gmail.com}
\affiliation{\mbox{Dipartimento di Fisica e Astronomia ``Galileo Galilei'', 
Universit\`a  di Padova, Via Marzolo 8, Padova 35131, Italy}}
\affiliation{ITAMP, Harvard-Smithsonian Center for Astrophysics, 60
  Garden Street, Cambridge, MA 02138, USA}
\date{\today}

\begin{abstract}
High precision spectroscopy of muonic hydrogen has recently led to
an anomaly in the Lamb shift, which has been parametrized in terms 
of a proton charge radius differing by seven standard deviations 
from the CODATA value.
We show how this anomaly may be explained, within about a factor of three, 
in the framework of an effective Yukawian gravitational potential
related to charged weak interactions, without additional free 
parameters with respect to the ones of the standard model.
The residual discrepancy from the experimental result in this model 
should be attributable to the approximations introduced in the 
calculation, the uncertainty in the exact value of the Fermi scale 
relevant to the model and the lack of detailed knowledge on
the gravitational radius of the proton. The latter cannot be inferred 
with electromagnetic probes due to the unknown gluonic contribution 
to the proton mass distribution. 
In this context, we argue that muonic hydrogen acts like a microscopic 
gravimeter suitable for testing a possible scenario for the reciprocal 
morphing between macroscopic gravitation and weak interactions, with the 
latter seen as the quantum, microscopic counterpart of the former.
\end{abstract}

\pacs{04.50.Kd, 04.60.Bc, 12.10.-g, 31.30.jr}

\maketitle

\section{Introduction}

Precision studies of the hydrogen spectroscopy have played a major role in shaping our 
knowledge of the microscopic world and its understanding in
terms of quantum mechanics and quantum field theory \cite{Reviewspectroscopy}. 
Precision spectroscopy of hydrogen has now reached a point where the accuracy in the comparison 
to quantum electrodynamics is limited by the proton size, in the form of 
the root-mean square (rms) charge radius, $r_p=\langle r_p^2 \rangle^{1/2}$. 
To test quantum electrodynamics at the highest precision level, the proton 
size should then be determined with high precision from independent experiments. 
In the analysis of electron scattering experiments, a value of
$r_p=(0.895 \pm 0.018)$ fm has been inferred \cite{Sick,Blunden},
while higher precision determinations are possible using muonic hydrogen \cite{Boriereview}. 
Since the more massive muon has a smaller Bohr radius and a more significant 
overlap with the proton, the correction due to the finite size of the latter 
is more significant than in usual hydrogen. 
However, a recent measurement \cite{Pohl} reported a value of $r_p=(0.84184\pm 0.00067)$ fm, which 
differs by seven standard deviations from the CODATA 2010 value of $(0.8775\pm 0.0051)$ fm, obtained 
by a combination of hydrogen spectroscopy and electron-proton scattering experiments. 
This anomaly corresponds to an excess of binding energy for the 2s state with respect 
to the 2p state equal to $\Delta E_{2s2p}^{\mathrm{exp}}=0.31$ meV. 
Since we expect no difference between electron and muons in their electromagnetic 
behavior, due to the underlying assumed lepton universality for electromagnetic interactions, 
this has resulted in what is called ``proton radius puzzle'' \cite{Pohlreview,Antogninireview}.

The existence of the proton radius puzzle seems confirmed by measurements of other energy 
levels allowing to determine the hyperfine structure with high precision \cite{Antognini}, 
and has recently generated a variety of theoretical hypotheses, including some invoking 
new degrees of freedom beyond the standard model \cite{Barger,Tucker}. 
In this context, some pioneering papers have already discussed high precision spectroscopy 
as a test of extra-dimensional physics  for hydrogen \cite{Luo1,Luo2}, helium-like ions \cite{Liu}, 
and muonium \cite{ZhiGang}, giving bounds on the number of extra-dimensions and their couplings. 
A recent attempt to explain the proton radius puzzle in the extra-dimensional setting has been
discussed in \cite{Wang}, and values for the coupling constant necessary to fit the anomaly were inferred. 

The idea that extra-dimensions may be in principle tested with atomic physics tools is appealing 
also considering the paucity ofviable experimental scenarios to test quantum gravity
\cite{Ellis,Amelino1,Amelino2,Amelino3}. 
However, it would be most compelling to have a setting in which this may be achieved in the most 
economic fashion, that is, without necessarily introducing new free parameters conveniently chosen 
to accommodate {\it a posteriori} the experimental facts. In this paper, we provide such an approach 
by exploring the consequences of a tentative unification between gravitation and weak interactions 
already conjectured in \cite{Onofrio}. In Section II, we introduce an effective potential energy 
between two pointlike masses which recovers Newtonian gravity at large distances, while morphing 
into an inverse square law interaction with strength equal to the one of weak charged
interactions at the Fermi scale, which coincides in our framework with the Planck scale. 
In Section III, we generalize this gravitational potential to the case of an extended structure 
like the proton, setting the stage for the evaluation of the Newtonian gravitational contribution to the
Lamb shifts in perturbation theory. Section IV contains the main result of this paper, that is, the 
evaluation of the Lamb shift for the Yukawian component.  The predicted contribution to the Lamb 
shift in muonic hydrogen is $\Delta E_{2s2p}=0.106$ meV, to be compared with the experimentally
determined value of $\Delta E_{2s2p}^{\mathrm{exp}}=0.31$ meV, {\it i.e.} a factor 2.8 discrepancy. 
One potential source of discrepancy between our prediction and the experimental result is then 
discussed more in detail. This is then followed by the predictions for the expected contribution
to the Lamb shift in muonic deuterium and a more qualitative discussion of the possible nature 
of the Yukawian potential of gravitoweak origin, including its selectivity towards the flavour 
of the fundamental fermions, this last feature being required in light of the manifest absence 
of a similar Lamb shift contribution for normal hydrogen. In Section V, we discuss possible 
tests in a purely leptonic system such as muonium, free from complications related to the extended
structure of hadrons. We show that the contribution of the gravitoweak Yukawian potential is 
negligible with respect to the current precision achieved in the experimental determination 
of observables such as the Lamb shift and higher precision observables such as the absolute 
1s2s transition frequency. 
In the conclusions, we stress that a more accurate evaluation calls for the measurement of the 
gravitational radius of the proton, which is expected to significantly differ from the charge radius 
due to the gluonic energy density distribution for which no experimental access seems available. 
A qualitative discussion of other systems in which the effective Yukawian potential introduced 
here may give rise to observable effects, or may give significant constraints on its parameters, 
concludes the paper.

\section{An effective potential for gravitation at short distances}

While we refer to \cite{Onofrio} for more details, we briefly recall here that the main 
idea we have explored is that what we call weak interactions, at least in their charged sector, 
should be considered as empirical manifestations of the quantized structure of gravity at or 
below the Fermi scale. This opens up a potential merging between weak interactions and gravity 
at the microscale, a possibility supported by earlier formal considerations on the physical 
consequences of the Einstein-Cartan theory \cite{Hehl}. Various attempts have been made in 
the past to introduce gravitoweak unifications schemes \cite{Hehl1,Batakis1,Batakis2,Loskutov}, 
and a possible running of the Newtonian gravitational constant in purely four-dimensional 
models has been recently discussed \cite{Capozziello,Calmet}. 
The conjecture discussed in \cite{Onofrio} relies upon identification of a quantitative 
relationship between the Fermi constant of weak interactions $G_F$ and a renormalized 
Newtonian universal gravitational constant $\tilde{G}_N$, that is, we may 
write\footnote{Equation 1 differs from Eq. (2) in \cite{Onofrio} since we have adopted 
in this paper a more rigorous definition of Planck mass as the one corresponding to the 
equality between the Compton wavelength and the Schwarzschild radius, {\it i.e.}, 
$\hbar/(M_P c)=2 G_N M_P/c^2$, the factor 2 in the Schwarzschild radius having been 
omitted in the first analysis presented in \cite{Onofrio}. Equality of the Planck energy 
and the vacuum expectation value of the Higgs field, $\tilde{E}_\mathrm{P}=v$, as in 
Eq. (4) of \cite{Onofrio} requires the new prefactor in Eq. (1) of this paper. 
The value of the Planck length is invariant as for $\tilde{E}_\mathrm{P}=v$ we have 
$\tilde{\Lambda}_\mathrm{P}=\hbar c/v$, regardless of the definition of the Planck mass.}
\begin{equation}
G_F=\sqrt{2}\left(\frac{\hbar}{c}\right)^2 \tilde{G}_N.
\end{equation}
This expression holds provided that we choose $\tilde{G}_N=1.229
\times 10^{33} G_N=8.205 \times 10^{22}~\mathrm{m^3~ kg^{-1}~ s^{-2}}$.
As an immediate benefit, the identification of the Fermi constant with a renormalized 
Newtonian universal constant via fundamental constants $\hbar$ and $c$ allow to identify 
Fermi and Planck scales as identical, $\tilde{E}_\mathrm{P}=v$, where $\tilde{E}_\mathrm{P}$ 
and $v$ are, respectively, the renormalized Planck energy and the vacuum expectation value 
of the Higgs field, sometimes called the Fermi scale, avoiding then any hierarchy issue.
As discussed in \cite{Onofrio}, there are  a number of possible tests of this conjecture
that can span a wide range of energies, from the ones involved in the search for gravitational-like 
forces below the millimeter range \cite{Onofrio1,Antoniadis}, to the ones explored at the Large 
Hadron Collider, with the spectroscopy of exotic atoms in between. 
Among the latter, we have outlined in \cite{Onofrio} the possibility that muonic hydrogen 
provides a suitable candidate, and here we make this proposition more concrete reporting 
an evaluation of this gravitational contribution to the Lamb shift in muonic hydrogen. 

\begin{figure}[t]
\includegraphics[width=0.60\columnwidth,clip]{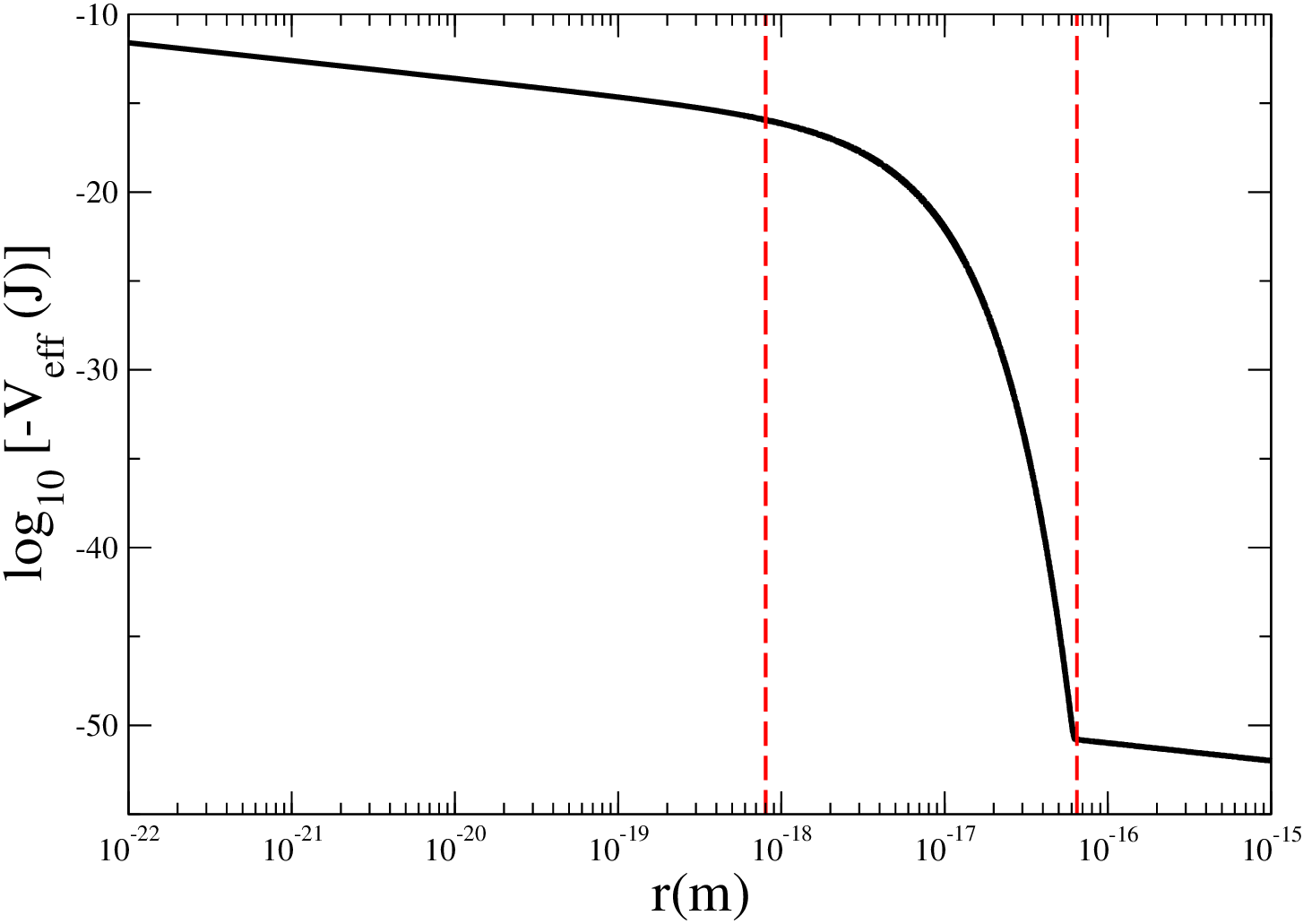}
\caption{Effective gravitational potential energy between an electron 
and a hypothetical pointlike proton versus their separation as from Eq. (2). 
The regions in which the inverse square law hold lie outside the two 
vertical dashed lines, as evidenced by the identical slopes.  In these regions, we expect 
degeneracy for the 2s-2p contribution within perturbation theory, while in between 
the dependence on distance removes the degeneracy resulting in a 
mass-dependent contribution to the Lamb shift. Notice that this
intermediate region extends over almost two decades starting at $r_\mathrm{min} \simeq
\tilde{\Lambda}_{\mathrm{P}}$ and ending at $r_\mathrm{max} \simeq
80 \tilde{\Lambda}_{\mathrm{P}}$, therefore boosting by almost two
orders of magnitude the influence of the lengthscale 
$\tilde{\Lambda}_{\mathrm{P}}$ at which quantum gravity effects 
are naturally expected to play a significant role.}
\label{muonichfig1}
\end{figure}

A first tool required for our analysis is a proper interpolation between the two regimes 
of weak gravity at macroscopic distances and the conjectured strong
gravity/weak interactions at the microscale. The existence of Newtonian gravitation at 
large distances, with coupling strength given by the universal gravitational 
constant $G_N$, which can morph into weak interactions corresponding to a 
renormalized universal gravitational constant $\tilde{G}_N$ at small distances, may be 
obtained by means of a generalized potential energy $V_\mathrm {eff}$
for the gravitational interaction between two pointlike particles of mass $m_1$ and $m_2$

\begin{equation}
V_\mathrm{eff}(r)=-\frac{G_N m_1 m_2}{r} \left(1+\alpha e^{-r/\lambda}\right)=
-\frac{G_N m_1 m_2}{r} \left[1+\left(\frac{\tilde{G}_N}{G_N}-1\right)
e^{-r/\tilde{\Lambda}_\mathrm{P}}\right],
\end{equation}
wherein the intermediate expression we have introduced, as customary in the analysis of 
Yukawian components of gravity \cite{Onofrio1,Antoniadis}, the generic parameters 
$\alpha$ and $\lambda$ for the strength and range of the Yukawian component respectively, 
thereby specialized in the last expression to our case of interest, 
$\alpha \equiv \tilde{G}_N/G_N-1$ and $\lambda \equiv \tilde{\Lambda}_\mathrm{P}$, with 
the renormalized Planck length
$\tilde{\Lambda}_\mathrm{P}=\sqrt{2\hbar \tilde{G}_N/c^3}=8.014 \times 10^{-19}$~ m. 

Equation 2 reduces to ordinary gravity for $r \gg \tilde{\Lambda}_\mathrm{P}$, 
whereas in the opposite regime of $r \ll \tilde{\Lambda}_\mathrm{P}$
continues to have a $1/r$ behaviour but with coupling strength proportional to 
$\tilde{G}_N$. In both limits, the evaluation in perturbation theory of the average 
gravitational energy should give no difference since $1/r$ potential is degenerate
for states with the same principal quantum number and different angular momenta. 
Therefore, the difference we may evidence in this analysis will reflect the genuine 
deviation from a inverse square law characteristic of Yukawian potentials. 
As manifest in Fig. 1 in the concrete example of an electron and a structureless, 
pointlike proton, the region in which the distance scaling of the corresponding 
force is not following the inverse square law extends over two decades starting 
at $r_\mathrm{min} \simeq  \tilde{\Lambda}_\mathrm{P}$.

\section{Gravitational energy between a lepton and an extended proton:
  Newtonian component}

The evaluation of the gravitational energy between two particles is quantitatively different 
if one of them has an extended structure, as in the case of the proton.
We schematize the proton as a spherical object with uniform mass ensity different from 
zero only within its electromagnetic radius $R_p$ related to the rms charge radius - 
for a uniform charge density - through $R_p=\sqrt{5/3}~r_p$, in such a way that 
 $\rho_p(r)=3m_p/(4\pi R_p^3)$ for $r\leq R_p$ and zero otherwise. 
With this assumption, the evaluation of the Newtonian potential energy - the first 
term in the righthand side  of Eq. (2) - between the proton and a generic lepton 
of mass $m_\ell$ ($\ell=e, \mu, \tau$) yields
 
\begin{equation}
V_{N\ell}(r) =\begin{cases} 
G_N \frac{m_\ell m_p r^2}{2R_p^3}-\frac{3}{2}G_N \frac{m_\ell m_p}{R_p}, & \hspace{1.0cm} 0<r<R_p, \\
- G_N \frac{m_\ell m_p}{r}, & \hspace{1.0cm} r>R_p. \
\end{cases}
\end{equation}
The calculation of the energy contribution due to the Newtonian potential may be performed 
by means of standard time-independent perturbation theory applied to 2s and 2p states. 
There is no space anisotropy, so we will focus on the radial (normalized) components of 
the unperturbed wavefunctions which are, respectively

\begin{equation}
R_{2s}(r) =  \frac{1}{(2a_\ell)^{3/2}}\left(2-\frac{r}{a_\ell}\right)
e^{-\frac{r}{2a_\ell}}; 
\hspace{1.0cm}
R_{2p}(r) = \frac{1}{\sqrt{3}(2a_\ell)^{3/2}}\frac{r}{a_\ell}e^{-\frac{r}{2a_\ell}},
\end{equation}
where $a_\ell$ is the Bohr radius, which also takes into account the reduced mass of the 
lepton-proton bound system. The calculation leads to
\begin{widetext}
\begin{eqnarray}
\langle V_{N\ell} \rangle_{2s}& = & \langle R_{2s}|V_{N\ell}|R_{2s}
 \rangle = -\frac{G_N m_\ell m_p}{R_p} \times \\ 
& & \left[-\frac{1}{16 \beta^2}\left(F_6\Big\vert_0^\beta-
4F_5\Big\vert_0^\beta+4F_4\Big\vert_0^\beta\right)
+\frac{3}{16}
\left(F_4\Big\vert_0^\beta-4F_3\Big\vert_0^\beta+4F_2\Big\vert_0^\beta\right)
+\frac{\beta}{8}\left(F_3\Big\vert_\beta^{+\infty}-4F_2\Big\vert_\beta^{+\infty}+
4F_1\Big\vert_\beta^{+\infty}\right)\right],\nonumber\\
\langle V_{N\ell} \rangle_{2p} & = &\langle R_{2p}|V_{N\ell}|R_{2p} \rangle =-\frac{G_N m_\ell m_p}{R_p} 
\left(-\frac{1}{48 \beta^2} F_6\Big\vert_0^\beta+\frac{1}{16} F_4\Big\vert_0^\beta
+ \frac{\beta}{24} F_3\Big\vert_\beta^{+\infty}\right), 
\end{eqnarray} 
\end{widetext}
where we have introduced $\beta=R_p/a_\ell$ and $F_n\Big\vert_a^b=\int_a^b dx x^n e^{-x}$.

\begin{figure}[t]
\includegraphics[width=0.60\columnwidth,clip]{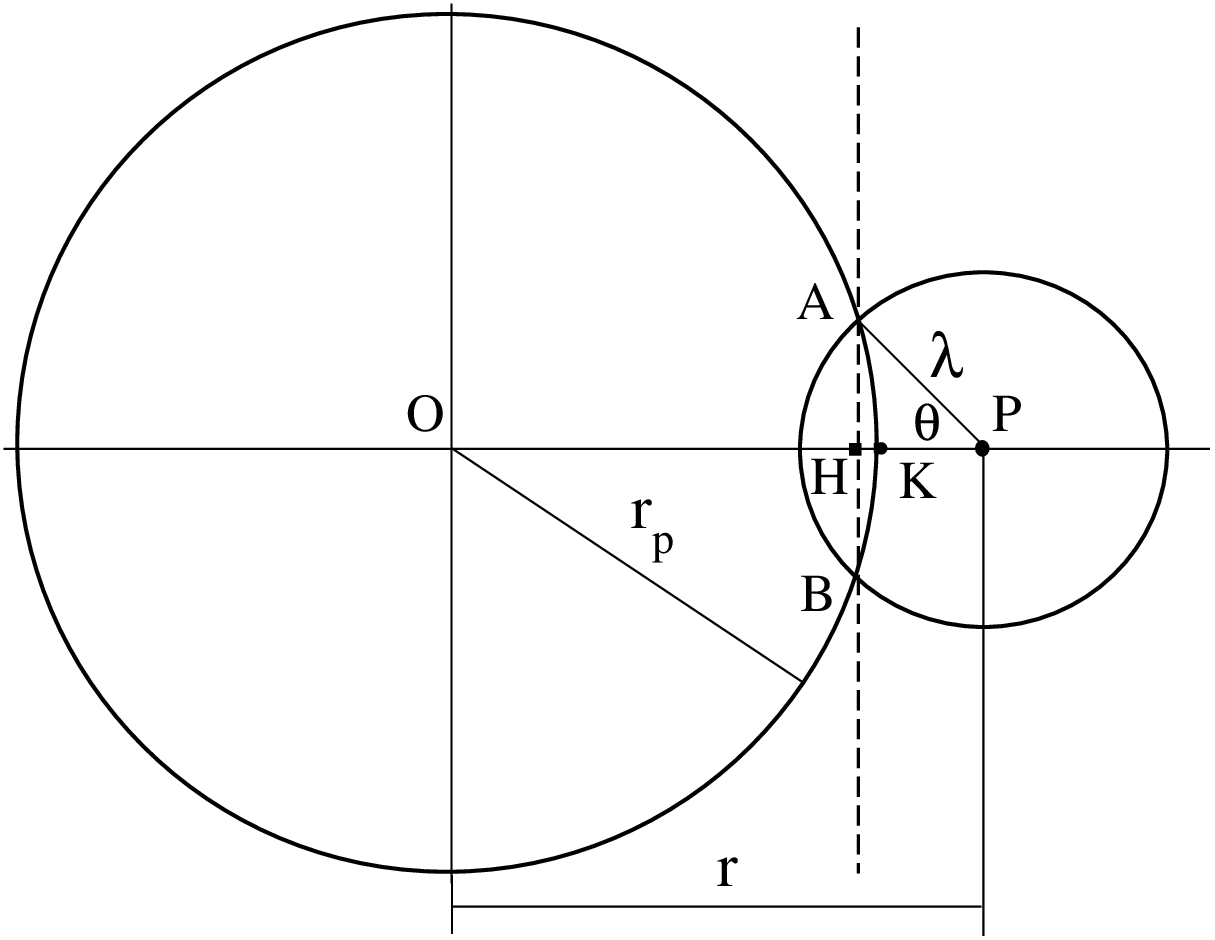}
\caption{Drawing (not to scale) of the relative proton and lepton coordinates for the 
evaluation of the Yukawian potential energy term due to the extended structure of the proton. 
The circle on the left side centered at point O represents the extent of the electric charge 
distribution of the proton, assumed uniform inside the sphere of radius $R_p$. 
The circle on the right side represents the range of the Yukawian interaction of 
a pointlike lepton located at point P.}
\label{muonichfig2}
\end{figure}

By straightforward algebraic manipulations, we obtain a more compact expression
for the Newtonian potential energy difference between the 2s and the 2p state as

\begin{equation}
\Delta \langle V_{N\ell} \rangle_{2s2p} \equiv \langle V_{N\ell} \rangle_{2s}-\langle
V_{N\ell} \rangle_{2p}=\frac{G_N m_\ell m_p}{R_p}\left[
6(e^\beta-1)\beta^{-2}-6\beta^{-1}-3-\beta-\frac{\beta^2}{4}\right]e^{-\beta},
\end{equation}
from which it is easy to see that, in the $\beta\rightarrow 0$ limit
of a pointlike proton, $\Delta \langle V_{N\ell}\rangle_{2s2p}\rightarrow 
G_N m_\ell m_p \beta^3 e^{-\beta}/(20 R_p)\rightarrow 0$. 
For a proton of finite size, this contribution is different for hydrogen and 
muonic hydrogen since the gravitational mass $m_\ell$ appears both as a factor 
and in the expression for the Bohr radius $a_\ell$ upon which $\beta$ depends. 
We notice also that this contribution tends to lift the energy of the 2s state, as this 
state is more sensitive to the finite size of the proton with respect to the 2p state, 
analogously to the case of the Coulombian attraction.
However, it is also easy to check that the difference is absolutely negligible in our 
context, being about 37 orders of magnitude smaller than the experimentally observed 
anomaly in muonic hydrogen, so it cannot play a role in its understanding. 
This is consistent with a simple estimate of the gravitational contribution in 
muonic hydrogen based upon consideration of a pointlike proton: from the estimate 
for the absolute Newtonian potential energy $V_g \simeq G_N m_{\mu} m_p/a_{\mu} 
\simeq 4.6 \times 10^{-34}$ eV, and obviously the presence of an extended 
structure for the proton, and the differential energy taken in the Lamb shift 
evaluation, add up to further suppress the Newtonian contribution. 
This also indicates however that if the Newtonian gravitational constant is 
boosted by 33 orders of magnitude as in replacing $G_N$ with $\tilde{G}_N$, 
the estimate for the absolute energy contribution is in the eV range, making 
its detailed evaluation relevant to the physics of muonic hydrogen.    

\begin{figure}[t]
\includegraphics[width=0.60\columnwidth,clip]{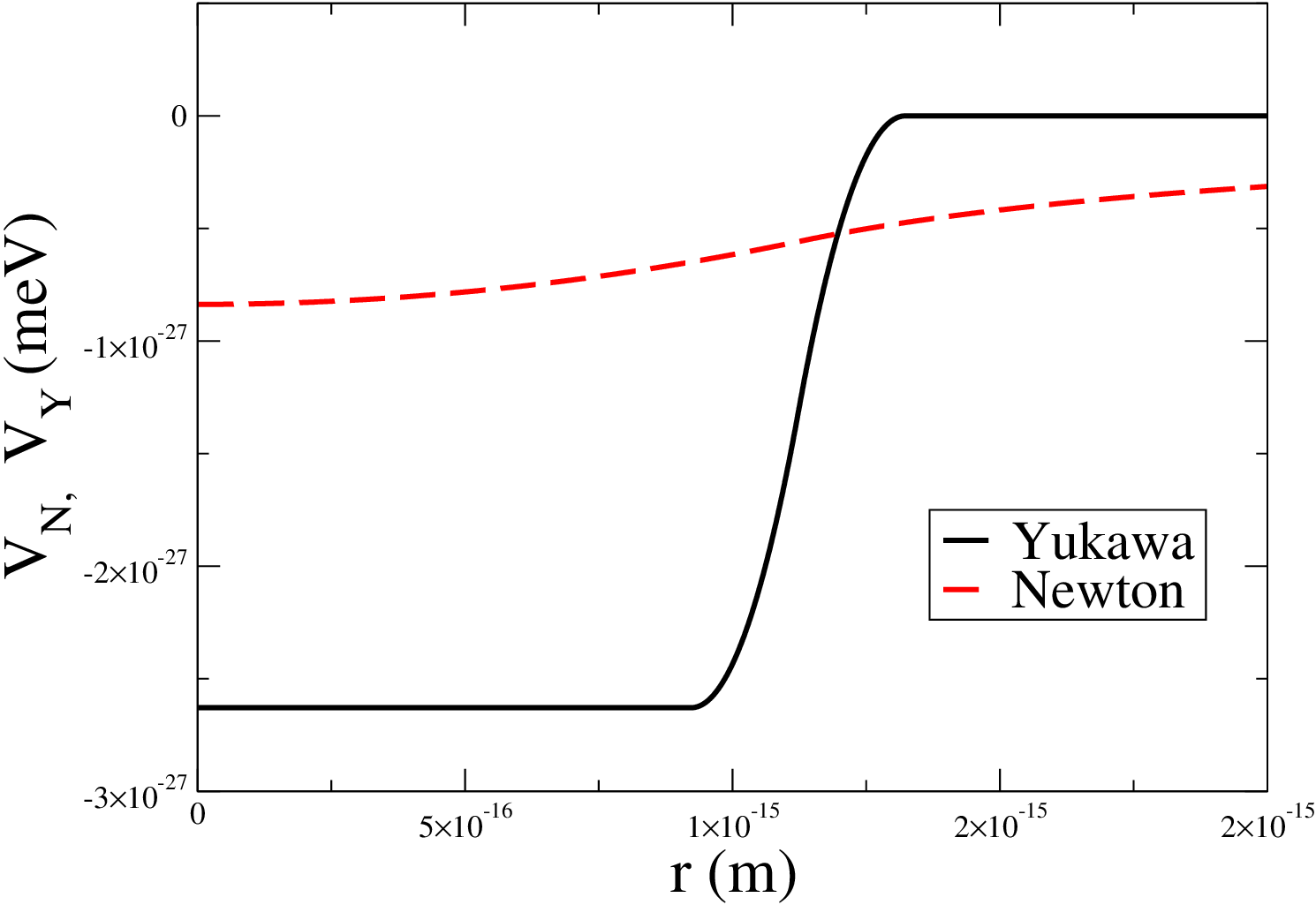}
\caption{Plot of the Yukawian (continuous black line) and Newtonian
  (dashed red line) gravitational potential energies in hydrogen 
corresponding to Eqs. (3) and (10), respectively.
The example assumes, to allow for an easier graphical comparison, 
interaction coupling strength and range as $\tilde{G}_N=10^2 G_N$, 
$\lambda=2.0\times 10^{-16}$ m, and rms charge radius $r_p=0.87 \times
  10^{-15}$ m, while the case considered in this paper has 
$\tilde{G}_N=1.229 \times 10^{33} G_N$ and $\lambda=\tilde{\Lambda}_\mathrm{P}=8.014 \times 10^{-19}$ m, leading
to a plateau value for the Yukawian component of the potential energy 
at small distance equal to $\simeq 0.52~ \mu$eV. 
The corresponding energy scale for muonic hydrogen is obtained by scaling 
up the vertical axis by the mass ratio of the two systems, leading to 
a plateau value $\simeq 0.108$ meV.} 
\label{muonichfig3}
\end{figure}

\section{Gravitational energy between a lepton and an extended proton:
Yukawian component}

The previous calculation sets the stage for the analysis of the Yukawian component, that is, 
the second term in the righthand side of Eq. (2). 
In a semiclassical picture, we expect that if the lepton is outside the
proton by at least an amount $\simeq \lambda$, there will
be a negligible Yukawian gravitational contribution. Likewise, once the lepton is 
completely inside the proton, there will be no Yukawian contribution either, as the 
uniform distribution of the proton mass will extert isotropic interactions. 
A nonzero value for the proton-lepton interaction occurs instead while the muon 
is partially penetrating inside the proton radius within a layer of order $\lambda$. 
The exact calculation for the Yukawian potential felt by the lepton should proceed by integrating 
the Yukawian potential contributions due to each infinitesimal volume element inside the proton. 
Since this calculation involves exponential integral functions and it cannot be trivially 
performed analytically,  we introduce two approximations. 
First, we truncate the Yukawian potential between two pointlike masses at distance $r$ in such a way that

\begin{equation}
V_{Y\ell}(r) = 
\begin{cases} 
- \alpha {G}_N \frac{m_\ell m_p}{r}, & \hspace{0.5cm}
0<r<\lambda, \\
0, & \hspace{0.5cm} r>\lambda.\
\end{cases}
\end{equation}

Therefore, the evaluation of the potential is carried out only in the region of intersection between 
two spheres, one of radius equal to $R_p$, the other of radius equal to the Yukawa range $\lambda$. 
In using this approximation, the amplitude of the Yukawian potential is then overestimated 
in the $0<r<\lambda$ region, while it is underestimated in the region corresponding to $r>\lambda$, 
as we will discuss more quantitatively at the end of the next section.
Second, we further assume that the intersection region is a spherical cap, an approximation 
corresponding to consider $R_p \gg \lambda$, which we will find {\it a posteriori} well satisfied.  

The potential felt by the lepton at a distance $r$ from the proton center is then evaluated by 
integrating the infinitesimal potential energy contributions over the volume of the spherical 
cap of the sphere of radius $\lambda$ centered on the lepton location. 
With reference to Fig. 2, and using spherical coordinates for the infinitesimal volume $dv$, 
this can be written as 

\begin{equation}
V_{Y\ell}(r)=-\alpha G_N m_\ell \rho_p \int \frac{dv}{\xi}=
-\alpha G_N m_\ell \rho_p \int_0^{2\pi} d\phi \int_0^{\bar{\theta}} dcos \theta
\int_{(r-R_p)/cos \theta}^{\lambda} \xi d\xi,   
\end{equation}
where $\cos \bar{\theta}=(r-R_p)/\lambda$. This leads to a Yukawian potential energy as follows

\begin{equation}
V_{Y\ell}(r) = 
\begin{cases}
- 2\pi \alpha G_N m_\ell \rho_p \lambda^2, & \hspace{1.0cm}  0<r<R_p-\lambda, \\
\pi \alpha G_N m_\ell \rho_p [(r-R_p)^2+2\lambda(r-R_p)-\lambda^2], & \hspace{1.0cm}  R_p-\lambda<r<R_p,\\
-  \pi \alpha G_N m_\ell \rho_p [(r-R_p)^2-2\lambda(r-R_p)+\lambda^2], & \hspace{1.0cm}  R_p<r<R_p+\lambda, \\
0, & \hspace{1.0cm} r>R_p+\lambda,\
\end{cases}
\end{equation}
in which continuity is ensured at all three boundary regions. As discussed above, this potential 
corresponds to a net attractive force only in the range $R_p-\lambda<r<R_p+\lambda$, and zero otherwise, 
and the plot of the corresponding potential energy is shown in Fig. 3.

We evaluate the expectation value of the Yukawian potential energy in the 2s and 2p states, 
$\langle V_{Y\ell} \rangle_{2s}$ and $\langle V_{Y\ell} \rangle_{2p}$ according to first-order 
perturbation theory. The Yukawian component of the gravitational potential is still much
smaller that the Coulombian potential even if it is coupled through $\tilde{G}_\mathrm{N}$ 
at short distances, since $\tilde{G}_\mathrm{N} m_\ell m_p <<e^2/(4\pi\epsilon_0)$. 
The calculation leads to

\begin{equation}
\begin{split}
\langle V_{Y\ell} \rangle_{2s} & =   - \frac{\pi \alpha G_N m_{\ell} \rho_p}{4}
\Big[\lambda^2 \left(F_4\Big\vert_0^y-4F_3\Big\vert_0^y+4F_2\Big\vert_0^y\right)
-\frac{a_\ell^2}{2} \left(F_6\Big\vert_y^\beta-2(2+v-\beta)F_5\Big\vert_y^\beta- \right. \\
& \left. (4-8v-8\beta+\beta^2-2\beta v-v^2) F_4\Big\vert_y^\beta-4 (2v-2\beta+\beta^2-2\beta v-v^2)
F_3\Big\vert_y^\beta+4(\beta^2-2\beta v-v^2)F_2\Big\vert_y^\beta \right) + \\ 
& + \frac{a_\ell^2}{2} \left(F_6\Big\vert_\beta^z-2(2+v+\beta)F_5\Big\vert_\beta^z+
(4+8v+8\beta+\beta^2+2\beta v+v^2)F_4 \Big\vert_\beta^z-4(2v+2\beta+\beta^2+2\beta
v+v^2)F_3\Big\vert_\beta^z + \right. \\  
& \left. + 4(\beta^2 +2 \beta v+v^2)F_2 \Big\vert_\beta^z \right)\Big],
\end{split}
\end{equation}

\begin{equation}
\begin{split}
\langle V_{Y\ell} \rangle_{2p} & =   - \frac{\pi \alpha G_N m_\ell \rho_p}{12} 
\Big[ \lambda^2 F_4 \Big\vert_0^y-\frac{a_\ell^2}{2} 
\left(F_6 \Big\vert_y^\beta+2(v-\beta) F_5\Big\vert_y^\beta+(\beta^2-2\beta
  v-v^2) F_4\Big\vert_y^\beta \right)+ \\
& + \frac{a_\ell^2}{2} \left(F_6 \Big\vert_\beta^z-2(v+\beta) F_5
\Big\vert_\beta^z+(\beta^2+2\beta v+v^2) F_4\Big\vert_\beta^z\right)\Big],
\end{split}
\end{equation} 
where $y=(R_p-\lambda)/a_\ell$, $z=(R_p+\lambda)/a_\ell$, and $v=\lambda/a_\ell$.

\begin{figure}[t]
\includegraphics[width=0.60\columnwidth,clip]{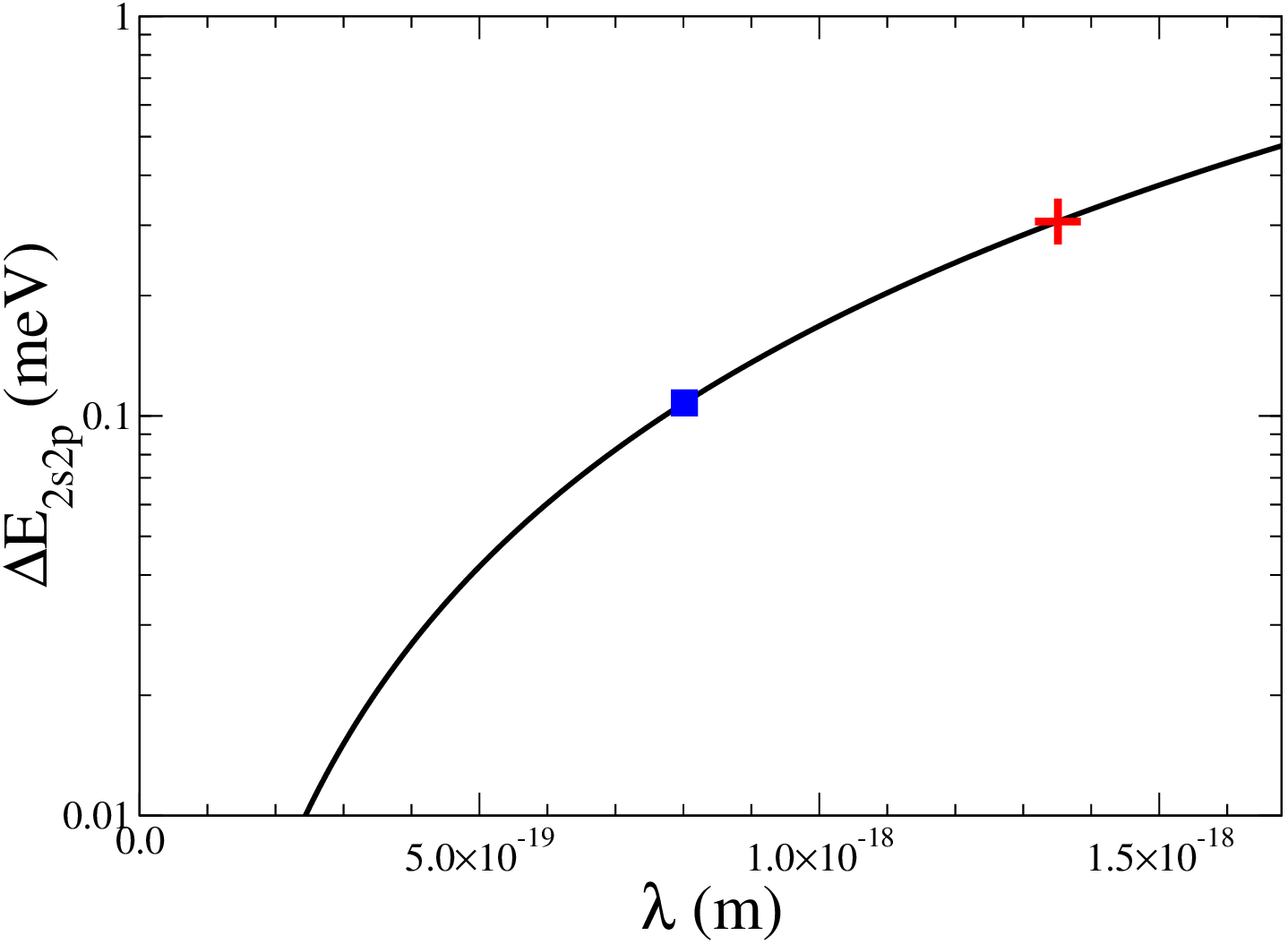}
\caption{Dependence of the calculated gravitational energy difference between the
2s and 2p energy levels of muonic hydrogen $\Delta E_{2s2p}$ versus the Yukawa range of 
short-distance gravity. 
We assume the effective coupling strength as $\tilde{G}_N=1.229 \times 10^{33} G_N=8.205 
\times 10^{22}~ \mathrm{m^3~ kg^{-1}~ s^{-2}}$ which gives rise to 
the Fermi coupling constant for charged weak interactions as discussed 
in \cite{Onofrio}. The CODATA 2010 value of the proton rms charge radius is assumed for the gravitational 
proton radius. The experimental value for $\Delta E_{2s2p}^{\mathrm{exp}}=$ 0.31
meV is also shown (red cross), which is explained assuming a value of $\lambda=1.35
\times 10^{-18}$ m at the effective coupling strength and proton radius assumed above. 
The prediction of the model for $\lambda=\tilde{\Lambda}_\mathrm{P}$, corresponding to
$\Delta E_{2s2p}$=0.106 meV, is marked by the blue square.}
\label{muonichfig4}
\end{figure}

The difference between the two contributions finally yields, after lengthy
algebraic simplifications, the expression
 
\begin{equation}
\Delta \langle V_{Y\ell} \rangle_{2s2p} \equiv
\langle V_Y \rangle_{2s}-\langle V_Y \rangle_{2p}=
-\frac{\alpha G_N m_\ell m_p}{16R_p^3}
\left\{\lambda^2 f(y)+a_\ell^2\left[g(y,\beta)+h(\beta,z)\right]\right\}, 
\end{equation}
where $f(y)=2(2-y)y^3 e^{-y}$ and

\begin{equation}
\begin{split}
g(y,\beta)
&=-G_6\Big\vert_y^\beta-2(v-\beta)G_5\Big\vert_y^\beta+2\left(-3+v-\beta+\frac{v^2}{2}+\beta
v-\frac{\beta^2}{2}\right)G_4\Big\vert_y^\beta+2(-12-2v+2\beta-v^2-2\beta 
v+\beta^2)G_3\Big\vert_y^\beta+
\nonumber \\
& 12(-1-v+\beta)G_2\Big\vert_y^\beta-24(6+v-\beta)\left(G_1\Big\vert_y^\beta+G_0\Big\vert_y^\beta\right),
\end{split}
\end{equation}

\begin{equation}
\begin{split}
h(\beta,z)
&=+G_6\Big\vert_\beta^z-2(v+\beta)G_5\Big\vert_\beta^z+2\left(3+v+\beta+\frac{v^2}{2}+\beta
v+\frac{\beta^2}{2}\right)G_4\Big\vert_\beta^z-2(-12+2v+2\beta+v^2+2\beta
v+\beta^2)G_3\Big\vert_\beta^z+
\nonumber \\
& 12(1-v-\beta)G_2\Big\vert_\beta^z+24(6-v-\beta)\left(G_1\Big\vert_\beta^z+G_0\Big\vert_\beta^z\right),\\
\end{split}
\end{equation}
where $G_n\Big\vert_a^b=a^n e^{-a}-b^n e^{-b}$.  The sign of the correction is negative, 
{\it i.e.} $\Delta E_{2s2p}<0$, indicating that the 2s state gets more bounded than the 2p state. 
This outcome is opposite to the case of the long-range Newtonian component in which
the 2s state gets a weaker binding since it explores more the proton interior where 
the net effect of gravity is smaller. This is due to the fact that the Yukawian potential 
does not fulfil the Gauss theorem and is marginally felt by the 2p state 
having small overlap with the proton region, unlike in the 2s state.
We will indicate from now on the absolute value of $\Delta E_{2s2p}$,
with the implicit understanding that it is negative.

The result of this analysis is shown in Figs. 4 and 5, which summarize the main findings of this paper. 
In Fig. 4, we show $\Delta E_{2s2p}$ as a function of the Yukawa range $\lambda$ for a fixed value of 
$\alpha$ corresponding to $\tilde{G}_N$ as assumed in the text immediately after Eq. (1). 
For a value of $\lambda=\tilde{\Lambda}_\mathrm{P}$, we obtain a value of $\Delta E_{2s2p} \simeq$ 
0.106 meV, about a factor 2.8 smaller than the measured value. 
The experimental value of $\Delta E_{2s2p}$ is instead obtained from Eq. (13) by assuming a 
value of $\lambda=1.35 \times 10^{-18} {\mathrm m} \simeq 1.7 \tilde{\Lambda}_\mathrm{P}$. 
In spite of the drastic assumption of uniform mass density for the proton inside its 
electric radius and of the two approximations used in the calculation, the proximity 
of the prediction on $\Delta E_{2s2p}$ assuming first-principle parameters $\alpha=\tilde{G}_N/G_N-1$ 
and $\lambda=\Tilde{\Lambda}_\mathrm{P}$ as inspired by the conjecture in \cite{Onofrio}, if not 
accidental, is rather remarkable.
Considering the uncertainty in the relevant proton radius, we have complemented the analysis by 
showing, in Fig. 5, the combined effect of $\lambda$ and $r_p$ on the evaluation of $\Delta E_{2s2p}$. 
We have also evaluated the points in the $(r_p,\lambda)$ plane which allow to obtain a value of 
$\Delta E_{2s2p}$ in the $(0.311,0.309)$ meV interval, see monotonically increasing line. 
The vertical band is the CODATA 2010 range of values for $r_p$, and the horizontal 
line corresponds to the value of $\lambda=\tilde{\Lambda}_\mathrm{P}$. 
To exactly verify our hypothesis we should have obtained a single intersection point, and 
this is not achieved within a relative accuracy conservatively estimated to be of order 
80 $\%$ in $\lambda$ and 40 $\%$ in $r_p$.

\begin{figure}[t]
\includegraphics[width=0.60\columnwidth,clip]{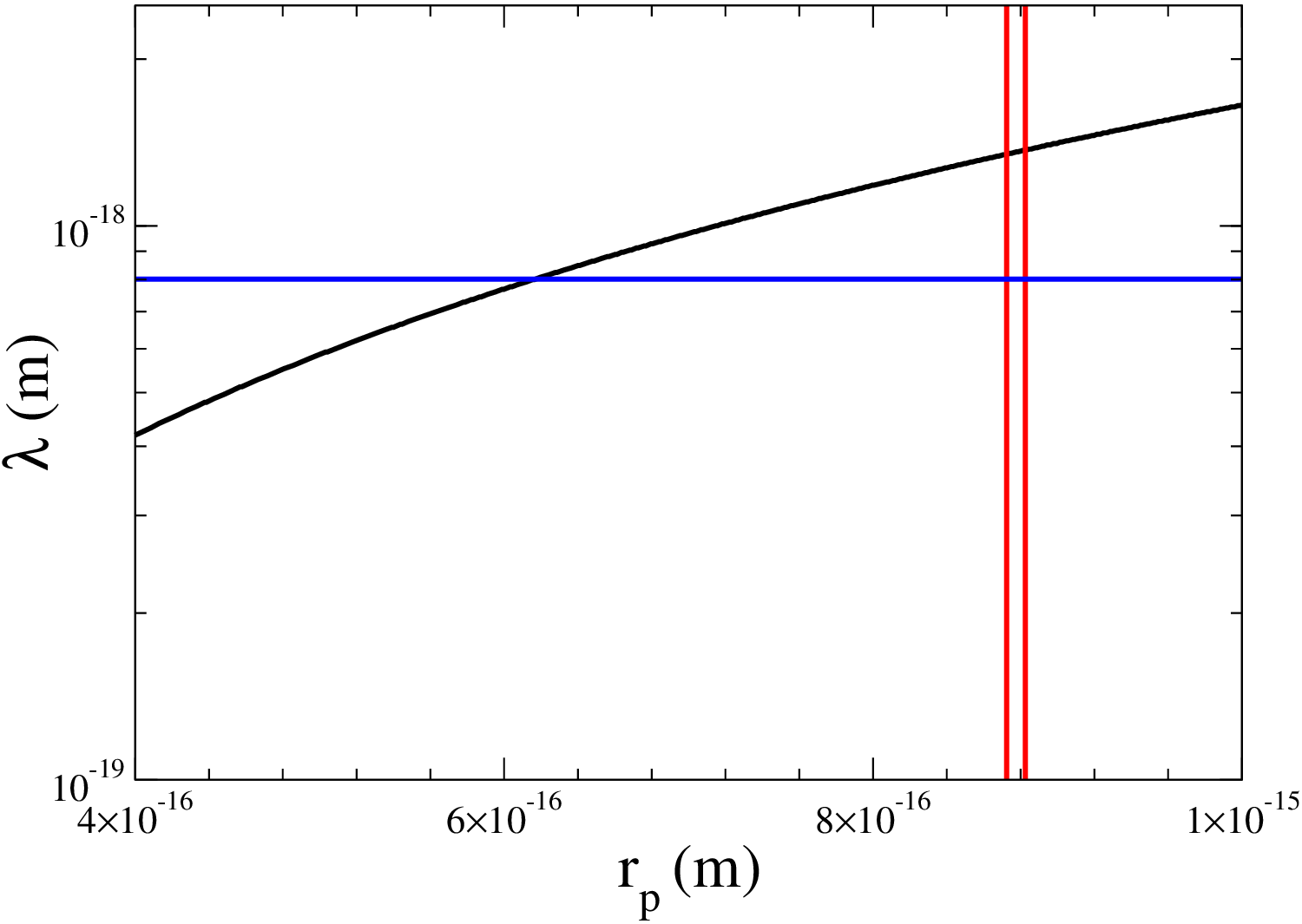}
\caption{Robustness of the Yukawian gravitational potential with respect to the Yukawa 
range $\lambda$ and the rms radius $r_p$. The locus of the points in the ($\lambda,r_p$) 
plane which give rise to a 0.309 meV $\leq\Delta E_{2s2p}\leq$ 0.311 meV is shown by the 
monotonically increasing curve. The range of values allowed, within one standard deviation, 
for the proton rms charge radius according to the CODATA value are delimited by the two vertical 
lines, the horizontal line corresponding to $\lambda=\tilde{\Lambda}_\mathrm{P}$. 
Complete agreement with the experimental value is obtained for instance by choosing 
$\lambda \simeq 1.7 \tilde{\Lambda}_\mathrm{P}$, while having $r_p$ at its CODATA value, or 
by choosing the gravitational radius of the proton equal to $\simeq$ 0.7 times the CODATA 
value for the rms charge radius while keeping $\lambda=\tilde{\Lambda}_\mathrm{P}$. }
\label{muonichfig5}
\end{figure}

We further notice that the second approximation, consisting in considering the intersection
region between the proton and the Yukawa sphere as a spherical cap, is well-satisfied since 
the proton radius is more than three orders of magnitude larger than the relevant Yukawa range. 
The truncation of the Yukawa potential expressed by Eq. (8) is instead a potential source of
discrepancy, which can be removed by a numerical analysis with the full Yukawa potential. 
While this is certainly a subject for future investigation, it is worth to point out that the 
truncated potential over distance does, in general, overestimate the integral of the Yukawa potential. 
This can be seen more quantitatively for pointlike particles by considering the zeroth moment of the
potential, its integral over the distance, $W_{Y}(r)= \int_r^{+\infty} d\xi \exp(-\xi/\lambda)/\xi$, and 
evaluating the coefficient $\sigma$ such that the zeroth moment of the truncated potential 
$W_{T\sigma}(r)=\int_r^{\sigma \lambda} d\xi /\xi$ equals $W_{Y}(r)$. 
Alternatively, one can evaluate a multiplicative coefficient $\gamma$ such that 
$W_{T\gamma}(r)= \gamma \int_r^{\lambda} d\xi/\xi$ equals $W_{Y}(r)$.
In Fig. 6 we show the dependence of these two correction parameters on
the distance $r$, evaluated in units of the Yukawa range $\lambda$. 
It is evident that in the region of small $r$ of interest for the evaluation of the 
integrals leading to the potential between the lepton and the proton
as an extended structure both $\sigma$ and $\gamma$ are smaller than one. 
This means that in substituting the Yukawa potential with its truncated 
approximation one should either replace $\Lambda_\mathrm{P}
\rightarrow \sigma \Lambda_\mathrm{P}$ or $\Delta
E_{2s2p}\rightarrow \gamma\Delta E_{2s2p}$. Both procedures go in
the direction of increasing the discrepancy discussed around 
Fig. 4, requiring a larger value of $\lambda$ or a smaller value of
$R_p$ to accommodate the experimental value of $\Delta E_{2s2p}$.

All possible refinements of the calculations are limited by the fact that in this 
approach the mass density distribution is crucial, and we do not have any information 
on the density distribution for the most important component of the proton at the 
level of determining its mass, {\it i.e.} the gluons. This seems currently the most critical 
issue preventing a more quantitative comparison of the model with the experimental result. 
It is plausible that gluons, only sensitive to the attractive color interaction, tend to 
cluster more than valence quarks which are also sensitive to the electromagnetic interaction 
acting both attractively (between the quarks up and down) and repulsively (between the two 
quarks up), thereby decreasing the effective gravitational radius of the proton below 
its electromagnetic value. 

In spite of the limitation arising from the lack of knowledge of the gravitational proton 
radius, it is worth to proceed with the analysis of other systems in which this putative 
Yukawian potential may also give rise to observable effects and predictions. 
In this context, here we limit our attention to muonic deuterium and hydrogen.  
In particular, on the one hand we are able to make a firm prediction for the Lamb shift 
excess to be expected in muonic deuterium.  
On the other hand, the analysis of Lamb shift in hydrogen shows that there
are potential troubles for the scenario envisaged here, which can be overcome only by
relaxing the demand for an economic approach with minimal assumptions. 

Muonic deuterium \cite{Carboni,Martynenko,Krutov} may be another viable test of our 
hypothesis since we expect a significant difference due to the deuterium mass and radius, 
apart from complications due to the internal mass distribution of deuterium. 
The calculation for muonic deuterium proceeds by introducing the deuteron charge radius 
\cite{Sick1} recommended by CODATA 2010 as $r_d=(2.1424 \pm 0.0021)$ fm, and obviously 
adjusting the mass as corresponding to the deuteron mass. 
We obtain, at $\lambda=\tilde{\Lambda}_\mathrm{P}$, an expected excess in the Lamb energy 
of 15.5 $\mu$eV, and of 42.6 $\mu$eV at $\lambda=1.35 \times 10^{-18}$ m at which the 
Lamb shift for muonic hydrogen is fully explained. This corresponds to an anomalous 
frequency shift in the 10 GHz range, smaller than the one in muonic hydrogen but still 
observable as well above the statistical (0.7 GHz) and systematic (0.3 GHz) uncertainties 
quoted in \cite{Pohl} for muonic hydrogen spectroscopy.

\begin{figure}[t]
\includegraphics[width=0.60\columnwidth,clip]{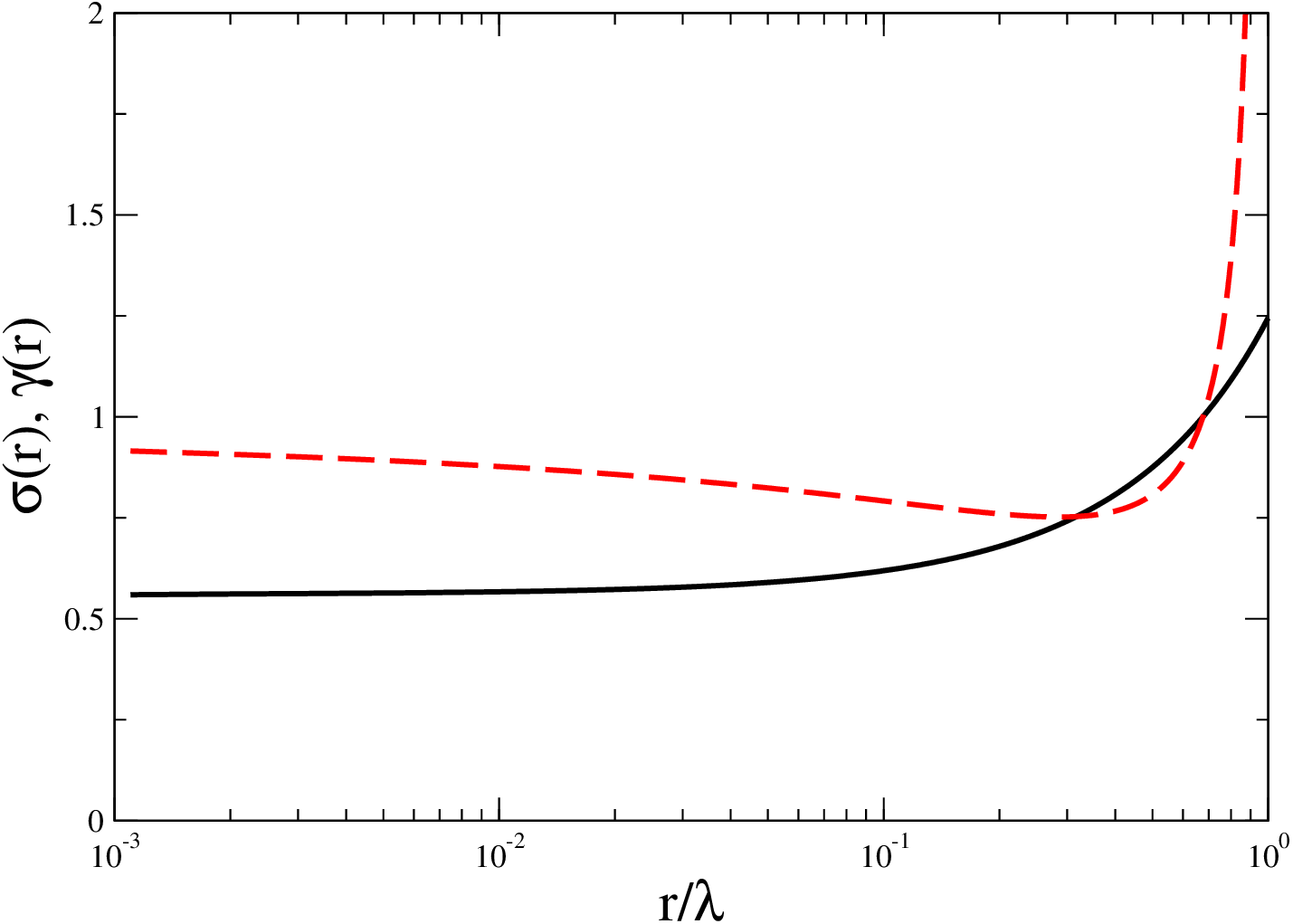}
\caption{Correction factors $\sigma$ (black, continuous line) and $\gamma$ (red, dashed line) 
for the truncated potential in Eq. (8) versus the lower integration bound for the zeroth order 
momentum of the potential energy between two pointlike particles. Both corrections confirm that 
the overall contribution of the truncated potential is overestimating the Yukawian potential energy.}
\label{muonichfig6}
\end{figure}

The predicted anomaly for the Lamb shift in (electronic) hydrogen is instead far too large, 
corresponding to 0.52 $\mu$eV at $\lambda=\tilde{\Lambda}_\mathrm{P}$ and 1.47 $\mu$eV at
$\lambda=1.35 \times 10^{-18}$ m, if using the same CODATA 2010 proton rms charge radius, and 
this is potentially a big issue for the validation of the proposed model. 
The Lamb shift in hydrogen is known with a precision which cannot incorporate such 
a large contribution, corresponding to an anomalous frequency shift of 0.2 GHz 
against an absolute accuracy of 9 kHz, or $8.5 \times 10^{-6}$ relative accuracy. 
Among possible ways to circumvent this issue - next to consider the whole model as invalidated - 
is to assume that the effective interaction corresponding to the Yukawian part in Eq. (2) is 
flavour-dependent, and does not act among fundamental fermions belonging to the same generation. 
This is not dissimilar from recent attempts \cite{Tucker,Barger1,Batell,Barger2} 
to introduce new interactions which differentiates between leptons, and may be  
also related to the need to understand the role of the electron-muon universality 
in the theory-experiment discrepancy for the anomalous magnetic moment of the muon 
\cite{Hertzog,Bennett,Jegerlehner}.
A putative ``off-diagonal'' interaction naturally spoils the universality characteristic 
of gravitation. However, it should be considered that in a possible gravitoweak unification 
scheme the emerging structure should presumably incorporate features of both weak and 
gravitational interactions. The former is manifestly flavor-dependent, as shown in the 
presence of Cabibbo-Kobayashi-Maskawa (CKM) and Pontecorvo-Maki-Nakagawa-Sakata (PMNS)  
mixing matrices for the charged current, so it is not {\it a priori} impossible that the 
interaction corresponding to the Yukawa component in Eq. (2) is highly selective in flavour content.
This solution could inspire searches for models in which a ``hidden sector'' of the standard model 
includes intermediate bosons of mass in the range of the Higgs vacuum expectation value 
mediating interactions which have mixed features between the usual charged weak interactions and 
gravitation, for instance heavier relatives of the $Z^0$ boson. 
As a final comment, we would like to also point out that {\it no} tests of the equivalence 
principle have ever been proposed involving other than the first generation of fundamental fermions. 
It would be interesting in this context to envisage tests of the equivalence principle, for instance 
free fall experiments involving strange matter such as neutral K mesons.

\section{Lamb-shift gravitational contribution in purely leptonic systems}

We finally complement the analysis by evaluating the contribution of the putative potential 
in Eq. (2) for purely leptonic systems, such as muonium, a $\mu^{\pm}e^{\mp}$ neutral bound state. 
Muonium is interesting from the theoretical viewpoint since, due to the pointlike structure of
leptons, there are no complications arising from a mass distribution as in the proton case, while 
experimentally there are complications due to the finite lifetime of the muon. 
Unfortunately, as we will see below, the expected effect is way too small to be observed, but 
the related calculations shed also light on the reason for which muonic hydrogen is so effective 
at spotting a Yukawa contribution with range in the attometer scale. 
The calculation of the perturbation to the energy levels for muons due to a potential like the 
one in Eq. (2) leads to the following expressions

\begin{eqnarray}
\langle V_{\mathrm{eff}} \rangle_{2s}&=&- \frac{G_N m_e m_{\mu}}{4a_{e\mu}} 
\left[1+\alpha \left(\frac{2}{(1+\xi)^2}-\frac{4}{(1+\xi)^3}+\frac{3}{(1+\xi)^4}\right)\right], \\
\langle V_{\mathrm{eff}} \rangle_{2p}&=&- \frac{G_n m_e m_{\mu}}{4a_{e\mu}} 
\left[1+\frac{\alpha}{(1+\xi)^4}\right], 
\end{eqnarray} 
where $\xi=a_{e\mu}/\lambda$. The contribution to the Lamb shift is then 

\begin{equation}
\Delta \langle V_{\mathrm{eff}} \rangle_{2s2p} \equiv
\langle V_{\mathrm{eff}} \rangle_{2s}-\langle V_{\mathrm{eff}} \rangle_{2p}=
- \alpha \frac{G_N m_e m_{\mu}}{4a_{e\mu}} 
\left[\frac{2}{(1+\xi)^2}-\frac{4}{(1+\xi)^3}+\frac{2}{(1+\xi)^4}\right], 
\end{equation}
in which the Newtonian $1/r$ contribution is obviously canceled out since it is 
equal in the 2s and 2p states. In the limit of $a_{e\mu} \gg \lambda$ of interest in 
our considerations, {\it i.e.} $\xi \gg 1$, the leading order term is the first in the 
right-hand side so we get a simplified, yet accurate expression as

\begin{equation}
\Delta \langle V_{\mathrm{eff}} \rangle_{2s2p} \simeq 
- \frac{\alpha G_N m_e m_{\mu} \lambda^2}{2a_{e\mu}^3}. 
\end{equation}
We obtain $\Delta \langle V_{\mathrm{eff}} \rangle_{2s2p} \simeq 1.85 \times 10^{-22}$ eV, corresponding 
to a frequency shift of $4.48 \times 10^{-8}$ Hz.  The precision in the determination of the Lamb 
shift in muonium, of order 1.5 $\%$ \cite{Oram,Badertscher,Woodle}, is not enough to observe the 
predicted effect in any foreseeable future. A precision observable in muonium is provided by the 
1s2s transition frequency. The contribution to the 1s state due to the effective gravitational potential is

\begin{equation}
\langle V_{\mathrm{eff}} \rangle_{1s}=
-\frac{G_N m_e m_{\mu}}{a_{e\mu}} \left[1+\frac{\alpha}{(1+\xi/2)^4}\right]. 
\end{equation}

The contribution to the 1s2s frequency then is 
 
\begin{equation}
\Delta \langle V_{\mathrm{eff}} \rangle_{1s2s} \simeq 
-\frac{7\alpha G_N m_e m_{\mu}\lambda^2}{2a_{e\mu}^3}, 
\end{equation}
which leads to a value of the same order of magnitude (just differing by a factor 7) of the 2s2p 
contribution evaluated above in Eq. (17). Although the 1s2s transition frequency is experimentally 
determined with a precision of 9.8 MHz \cite{Meyer}, the expected signal is still too low to provide 
an observable frequency shift. The situation in principle is less pessimistic for a $\mu^+\mu^-$ state 
(the so-called ``true muonium''), as the Lamb shift and the 1s2s contributions get increased by a factor 
$(m_\mu/m_e)^4/8$, leading to $\simeq$ 10 Hz for the 2s2p frequency shift, but the very experimental 
feasibility of muonium is still under debate \cite{Brodsky}, due to the complications of producing 
a bound state of two, rather than one, unstable particles. 

Nevertheless, this exercise shows the efficacy of muonic hydrogen with respect to purely leptonic, 
structureless, systems, in which the Bohr radius and the Yukawa range are the only relevant lengthscales. 
The presence of an intermediate lengthscale, the proton radius, gives a boost in the expected 
contribution of muonic hydrogen with respect to its muonium counterpart by an amount equal to 
$(a_{\mu}/R_p)^3 \simeq 8 \times 10^6$, as arguable by means of simple dimensional considerations.
In other words, while leptonic systems get a contribution which is quite far from experimental 
observation, semileptonic systems such as muonic hydrogen instead get an amplified contribution 
due to the extended structure of the proton which shifts down the relevant lengthscale from $a_\ell$
to $R_p$, although at the price of introducing a parameter, the proton radius, which is not yet 
known from first principles.
This presumably shows the existence of a narrow window for observing
effects due to quantum gravity at the attometer (Fermi) scale.  
In any event, the fact that muonic hydrogen discriminates between 2s and 2p 
states, with respect to a Yukawian component of strength much larger than Newtonian 
gravity and acting on the attometer scale, should make it the first effective ``quantum gravimeter''.   

\section{Conclusions}

In conclusion, we have shown that a solution to the ``proton radius puzzle'' is potentially 
available by means of an effective Yukawian potential originated by the morphing of Newtonian 
gravitation into weak interactions at the Fermi scale as conjectured in \cite{Onofrio}, without 
additional parameters with respect to the ones already present in the standard model.  
Leaving aside the need for removing the approximations used in the evaluation of the 
Yukawa potential due to the proton as an extended object, either the relevant Yukawa 
range is not coinciding with the Planck scale, or the relevant proton radius is not the 
one determined using electric probes, or a combination of both factors. 
Among all these considerations, it is important to point out that the electric charge distribution of 
the proton investigated by using leptonic probes is not accurately representative of the mass distribution. 
Within the standard model gluons - which account for a large fraction of the proton mass - do not 
have electroweak interactions and therefore charged lepton or neutrino scattering off protons 
cannot provide any information on their spatial distribution. 
Furthermore, we have been forced to assume that the electron does not interact with the proton 
via the same effective Yukawian interaction as the muon, since the expected anomalous contribution 
in the former case is exceedingly large with respect to what is observed in the hydrogen Lamb shift. 
Although this spoils the requirement for simplicity and universality, it is plausible in 
light of the complex structure of weak interactions which are flavour- and generation-dependent. 

Lastly, it may be worth to briefly discuss possible consequences of our model for future experimental 
investigations. A very important proposal, as the one described in \cite{Gilman} on muon-proton 
scattering, will not add insights on this issue, yet it will be critical in corroborating the existence 
of the anomaly and to pinpoint any possible deviation from the electron-muon universality. 
Unfortunately, the intrinsically short lifetime of the $\tau$-lepton of 291 fs makes it 
unfeasible to perform spectroscopic measurements on tauonic hydrogen, for which the predicted 
gravitational anomalous Lamb shift is even larger than that for muonic hydrogen. 
Also, provided that strong interaction effects can be effectively filtered out at the 
required accuracy level, a systematic study of precision spectroscopy of other exotic atoms 
with Bohr radii below the picometer range (see \cite{Nesvi} for a pioneering analysis), such 
as protonium \cite{Venturelli,Lodi}, may allow to confirm or disproof our model in a reasonable time-frame.

\acknowledgments
It is a pleasure to thank Lorenza Viola for a critical reading of the manuscript.

\end{document}